\documentclass[twocolumn,showpacs,preprintnumbers,amsmath,amssymb,aps]{revtex4}

\newcommand{\titel}{Global stability criterion for
a quantum feedback control process on a single qubit
and exponential stability in case of perfect detection efficiency}
\newcommand{\fett}[1]{\mbox{\mathversion{bold}$#1$}}

\usepackage{mathrsfs}
\usepackage{color}
\usepackage[latin1]{inputenc} 

\newcommand{\D}{\,\mathrm{d}}
\newcommand{\E}{\,\mathrm{e}}

\newcommand{\I}{\mathrm{i}}

\usepackage{ifpdf}
\ifpdf
  \usepackage[pdftex]{graphicx}
  \usepackage[pdftitle={\titel},
              pdfauthor={Andreas de Vries},%
              pdfpagemode=FirstPage,%
              colorlinks=false,hyperindex=true,plainpages=false,%
  ]{hyperref}
  \DeclareGraphicsExtensions{.pdf,.jpg,.png}
\else
  \usepackage{graphicx}
  \usepackage[pdftitle={\titel},
              pdfauthor={Andreas de Vries},%
              pdfpagemode=FirstPage,%
              colorlinks=false,hyperindex=true,plainpages=false,%
              dvips=true,ps2pdf=true%
  ]{hyperref}
  \DeclareGraphicsExtensions{.eps}
\fi

\newcommand{\gruen}[1]{\textcolor[rgb]{0,.6,0}   {\textbf{#1}}}

\newtheorem{satz}{Theorem}

\newtheorem{defi}[satz]{Definition}

\newtheorem{axi}[satz]{Axiom}

\newtheorem{rem}[satz]{Remark}
\newtheorem{beisp}[satz]{Example}
\newtheorem{beispe}[satz]{Examples}

\newenvironment{beweis}{\emph{Proof.}}{\hfill{$\square$}\\ }
\newenvironment{definition} {\begin{defi}\begin{em}}{\end{em}\hfill{$\square$}\end{defi}}
\newenvironment{beispiel} {\begin{beisp}\begin{em}}{\end{em}\hfill{$\square$}\end{beisp}}

\begin{document}

\title{\titel}

\author{Andreas de Vries}%
  \email{de-vries@fh-swf.de}
\affiliation{%
		FH S\"udwestfalen University of Applied Sciences, 
		Haldener Stra{\ss}e 182, 
		D-58095 Hagen,
		Germany%
}%
\altaffiliation[Also at ]{AG Mathem.\ Physik, Ruhr-Universit\"at, D-44780 Bochum}


\date{\today}


\begin{abstract}
	Quantum feedback control is a technology which can be used
	to drive a quantum system into a predetermined eigenstate.
	In this article, sufficient conditions for the experiment
	parameters of a quantum feedback control process of a homodyne QND measurement 
	are given to guarantee feedback control of a
	spin-1/2 quantum system
	in case of imperfect detection efficiency.
	For the case of pure states and perfect detection efficiency,
	time scales of feedback control processes are calculated.
\end{abstract}

\maketitle


\section{Introduction}
In classical control theory, feedback control describes processes in which
a closed-loop controller is used to steer the states or outputs of a dynamical system,
which in turn effect the inputs of the controller into the system.
A remarkable approach to feedback control of quantum spin systems has recently 
been elaborated in 
\cite{van-Handel-et-al-2005}.
Here QND measurements are utilized to let a quantum system collapse
deterministically onto a predetermined eigenstate.

In the present article, the stability and
the time scale of quantum feedback control processes are studied.
As a result (Theorem \ref{satz-asymptotically-stable}),
sufficient limits for the experiment
control parameters are derived to guarantee asymptotically stable 
quantum feedback control processes
on a spin-$\frac12$ quantum system.
It is proved by applying Lyapunov's method to the stochastic
differential equation governing the quantum state evolution,
and thus differs from the 
similar result in \cite{van-Handel-et-al-2005} proposing numerical methods
of semialgebraic geometry and aiming at applicability for higher
spin systems where efficient search for Lyapunov functions is practically impossible.
For the special case of pure states and perfect detection efficiency, the quantum feedback control process
is proved to terminate even exponentially fast in time.

The article is organized as follows. First,
the notions of QND measurements and
quantum feedback control
for a spin-$\frac12$ system
are shortly reviewed, before
the stochastic stability of quantum feedback control processes 
with imperfect and perfect detection efficiency are studied,
and the results are shortly discussed.

\section{QND measurements}
In contrast to a measurement in classical physics, a quantum measurement inevitably
changes, or even destroys, the measured quantum system itself
\cite{Goswami-1997,Nielsen-Chuang-2000}.
Theoretical as well as experimental investigation have been intensively made
dealing with processes where quantum measurements are utilized constructively,
for instance
theoretical considerations of measurement determination
by the quantum register \cite{Dusek-Buzek-2002},
quantum feedback control by continuous measurements 
\cite{Belavkin-1992,Belavkin-1992b,Belavkin-1994,Wiseman-1994},
especially in quantum optics 
\cite{Armen-et-al-2002, Stockton-et-al-2002, Stockton-et-al-2004, van-Handel-et-al-2005},
stabilization and purification of two-level systems
\cite{Wiseman-et-al-2002,Wiseman-Ralph-2006},
conditional measurements of coupled quantum dots by a point contact detector
\cite{Goan-Milburn-2001,Fujisawa-et-al-2004}
or by a SET \cite{Shnirman-Schoen-1998,Gurvitz-2003,Gurvitz-Berman-2005},
and the conditional measurement
approach due to Sherman and Kurizki \cite{Sherman-Kurizki-1992}
to prepare predetermined field states of atoms trapped in optical QED cavities
\cite{Harel-et-al-1996,Fortunato-et-al-1996,Fortunato-et-al-1999},
as well as a similar approach analyzed for spin squeezing in Cs clocks
\cite{Oblak-et-al-2005}.

Although these approaches differ considerably in detail,
most of them utilize 
repeated quantum nondemolition (QND) measurements
\cite[§3.3]{Joos-et-al-2003}, i.e.,
measurements of an observable $Y$ 
satisfying
the \emph{self-nondemolition condition}
$ 
	[Y(t), Y(t')] = 0
$ 
for all times $t$, $t'$, as well as
the \emph{back action evasion condition}
$ 
	[Y,H_{\mathrm{int}}] = 0,
$ 
where 
$H_{\mathrm{int}} = \sum_j |j\rangle \langle j| \otimes B_j$ 
denotes the interaction Hamiltonian between the
considered system (the projections $|j\rangle \langle j|$)
and the measuring apparatus ($B_j$).
The QND observable $Y$ may correspond, for instance, to a Hermitian 
Lindblad operator $L$, or to a conserved quantity, 
such as a constant of motion of the considered system
like  polarization or momentum. 

\section{Quantum feedback control}
Due to ideas of 
Belavkin \cite{Belavkin-1992,Belavkin-1992b,Belavkin-1994}
as well as Wiseman and coworkers
\cite{Wiseman-1994,Thomsen-et-al-2002}, 
an approach to quantum feedback control
of spin systems has been recently developed by van Handel, Stockton, and Mabuchi
\cite{van-Handel-et-al-2005}.
In this approach repeated quantum nondemolition measurements are 
engineered to let quantum spin systems collapse
deterministically onto a previously chosen eigenstate.
This idea, surprising from a traditional physics perspective,
bases on the fact that realistic measurements are not instantaneous
but take some finite time. If these reduction time scales are
of an order attainable by modern digital electronics, 
a quantum filter and a controller can respond on the spin system state,
feeding back the intermediate 
nondemolition measurement results to a Hamiltonian parameter.
In this way it is possible, for instance, to deterministically prepare
highly entangled Dicke states \cite{Stockton-et-al-2004},
to generate and
utilize squeezed quantum states of trapped atoms 
in an optical cavity \cite{Oblak-et-al-2005},
or to improve quantum error correction \cite{Ahn-et-al-2002}.

\begin{figure}[h!tp]
	\centering
	\begin{footnotesize}
	\unitlength1ex
	\begin{picture}(49,16)
		\put( 0,-0.3){\includegraphics[width=50ex]{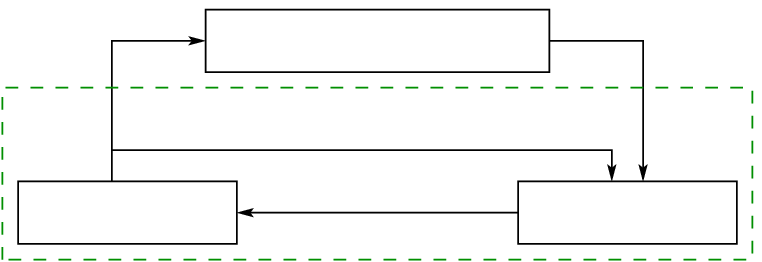}}
		\put(25.0,14.3){\makebox(0,0){Quantum system}}
		\put(40.0,14.5){\makebox(0,0)[b]{$y_t$}}
		\put(25.0, 9.5){\makebox(0,0){\gruen{\em Digital processing}}}
		\put( 7.0, 7.5){\makebox(0,0)[r]{$H(t)$}}
		\put( 8.5, 3.2){\makebox(0,0){Controller}}
		\put(25.0, 4.0){\makebox(0,0){$\rho_t$}}
		\put(41.5, 3.2){\makebox(0,0){Filter}}
	\end{picture}
	\end{footnotesize}
	\caption{\label{fig-control}
		(Color online)
		Schema of a quantum feedback control process.
		The QND measurement output $y_t$ from the quantum system
		is used to propagate the conditional state of the 
		filter, via the feedback signal $H(t)$.
		The dashed line indicates (classical) 
		digital processing, the filter is determined by
		Eq.~(\ref{eq-SME}).
	}
\end{figure}

The quantum stochastic control formalism of van Handel, Stockton, and Mabuchi
\cite{van-Handel-et-al-2005}
can be considered as an extension of probability theory,
and the traditional formulation
of quantum mechanics can be directly recovered from it.
In \cite[§IV.C]{van-Handel-et-al-2005} a stabilizing controller
is given for a quantum system of spin $j=\frac12$,
schematically depicted in Figure \ref{fig-control}.
The 
conditional evolution
of the density operator $\rho$ describing the quantum system 
depends on the probe parameter measurement rate $M>0$ in Hz, and
the detection efficiency $\eta$ $\in$ $[0,1]$, a pure number.
More precisely, the conditional evolution of $\rho$
is determined by the stochastic master equation
\begin{equation}
	\D\rho_t
	=
	\mathscr{G}^*[H(t), L] \rho_t \D t
	+\sqrt\eta\, \mathscr{H}[L] \rho_t
	\D W_t
	,
	\label{eq-SME}
\end{equation}
where $H(t)$ is the control Hamiltonian (with $H(t)=0$ in case of no feedback),
$L$ is an observable 
one of whose eigenstates is the desired final state of the system,
$\mathscr{G}^* = \mathscr{G}^*[H(t),L]$ is the adjoint generator
\begin{equation}
	\mathscr{G}^* \rho_t = -\I [H(t), \rho_t] 
	+
	L \rho_t L^* - {\textstyle \frac12} (L^* L \rho_t + \rho_t L^* L)
	,
\end{equation}
$\mathscr{H}$ is the superoperator
\begin{eqnarray}
	\mathscr{H}[L] \rho_t
	& \hspace*{-1.0ex} = \hspace*{-1.0ex} &
	L \rho_t + \rho_t L^* - \mathrm{Tr}[\rho_t (L+L^*)]\,\rho_t,
\end{eqnarray}
and $\D W_t$ is the innovations process
\begin{equation}
	\D W_t = 
	2 \sqrt{M\eta}\, y_t \D t 
	- \sqrt\eta \, \mathrm{Tr}[\rho_t (L+L^*)]\,  \D t
\end{equation}	
depending on 
the QND measurement record $y_t$ of the output corresponding to the observable $Y(t)$
\cite[§II]{Stockton-et-al-2004}.
Here $Y$ is normalized (i.e., $\D Y_t^2 = \D t$) and
related to $L$ and the standard noises $A$ and $A^*$ 
by the Hudson-Parthasarathy equation.
The innovations $\D W_t$ is a Wiener increment and $\D W_t/\D t$ is a
Gaussian white noise. Note that $\D W_t$ is one-dimensional, whereas
Eq.~(\ref{eq-SME}) is operator-valued.

Usually, the controller Hamiltonian $H(t)$ is determined by a few
control parameters.
It is most desirable if they 
could be adjusted in a way such that the master equation
has an asymptotically stable fixed point.

\subsection{Density operator space of a spin-$\frac12$ system}
In the prototypical physical model
of homodyne measurement of a spin system
\cite{van-Handel-et-al-2005},
where $y_t$ denotes the homodyne measurement record of the output,
the observable $L$ and the controller Hamiltonian $H(t)$ in Eq.~(\ref{eq-SME})
are given by
\begin{equation}
	L=\sqrt{M} J_z
	\qquad
	\mbox{and}
	\qquad
	H(t)=B(t) J_y
\end{equation}
with the usual angular momentum observables $J_y$, $J_z$,
and $\D W_t/\D t$ can be identified with the shot-noise of the
homodyne local oscillator.
Here $M > 0$ is the strength of 
the interaction between the light and the atoms and is regulated 
experimentally by the optical cavity, 
and the control input $B(t)$ is the applied magnetic field.
The time scale 
then only depends on the sensitivity $1/(2\sqrt{M\eta})$ of the photodetection
per $\sqrt{\mathrm{Hz}}$,
and the feedback gain parameters 
of the controller $B(t)$.
Hence the time scale only depends on experimentally controlled parameters.
For perfect detection efficiency, $\eta=1$, the stochastic master equation
governing a pure quantum system under feedback control is one-dimensional
and will be tackled analytically below.

For the special case 
of a quantum system of spin $\frac12$, i.e., a qubit,
the space of all density operators $\rho$ of the system is represented by
the two-dimensional disc
\begin{equation}
	D^2
	= \{ (\lambda, \nu) \in \mathbb{R}^2: \lambda^2 + \nu(\nu-1) \leqq 0\}
\end{equation}
with center $(0,\frac12)$ and radius $\frac12$,
\cite[§IV.C]{van-Handel-et-al-2005}.
Here the density matrix entries are given by 
$\rho_{11} = \nu$, $\rho_{22} = 1 - \nu$, $\rho_{21} = \rho_{12}^* = \lambda$,
and the state $(\lambda,\nu) = (0,0)$ to be stabilized corresponds to
$\rho$ $=$ $|1\rangle \langle 1|$ $=$ diag$\,(0,1)$. 
\begin{figure}[h!tp]
	\centering
	\begin{footnotesize}
	\unitlength1ex
	\begin{picture}(20,16.5)
		\put( 0,-0.5){\includegraphics[width=20ex]{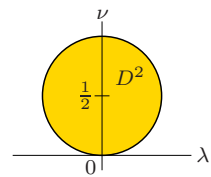}}
		\put(10,16.3){\makebox(0,0)[b]{$\nu$}}
		\put(13.0,10.0){\makebox(0,0){$D^2$}}
		\put(9.0,7.8){\makebox(0,0)[r]{$\frac12$}}
		\put(9.3,0.0){\makebox(0,0)[ur]{$0$}}
		\put(20.3,1.3){\makebox(0,0)[l]{$\lambda$}}
	\end{picture}
	\end{footnotesize}
	\caption{\label{fig-D-2} 
		(Color online)
		The density operator space $D^2$ of a spin-$\frac12$ quantum system. 
		The origin $(\lambda, \nu) = (0,0)$
		corresponds to the quantum state $|1\rangle \langle 1|$, 
		the point $(0,1)$
		to $|0\rangle \langle 0|$.
	}
\end{figure}
Note that in this case the imaginary
parts of the off-diagonal entries of $\rho$ decouple
and may be neglected.
The stochastic master equation (\ref{eq-SME})
describing the conditional evolution of a single qubit thus is
reducible to the two-dimensional Itô equation \cite{van-Handel-et-al-2005}
\begin{eqnarray}
	\D\lambda_t
	& \hspace*{-1.0ex} = \hspace*{-1.0ex} & \textstyle
	[B(t) (\nu_t - \frac12) - \frac M2 \lambda_t] \D t
	+ \sqrt{M\eta}\, \lambda_t (1-2\nu_t)\D W_t
	,
	\nonumber \\
	\D\nu_t
	& \hspace*{-1.0ex} = \hspace*{-1.0ex} & \textstyle
	 - B(t) \lambda_t \D t
	 - 2 \sqrt{M\eta}\, \nu_t (\nu_t-1)\D W_t
	.
	\label{eq-SME-j=1/2}
\end{eqnarray}
Its infinitesimal generator \cite[§7.3]{Oeksendal-1998} is given as
\begin{eqnarray}
	\mathscr{L} 
	& \hspace*{-1.0ex} = \hspace*{-1.0ex} & \textstyle
	[B(\lambda,\nu) (\nu - \frac12) 
	- \frac M2 \lambda] \frac{\partial}{\partial \lambda}
	- B(\lambda,\nu) \lambda \frac{\partial}{\partial \nu}
	\nonumber \\*
	& \hspace*{-1.0ex} \hspace*{-1.0ex} & \textstyle
	{}+ 2M\eta \left[\lambda^2 (\nu - \frac12)^2
	\frac{\partial^2}{\partial \lambda^2}
	+ \nu^2 (\nu - 1)^2 
	\frac{\partial^2}{\partial \nu^2}
	\right]
	,
\end{eqnarray}
so $\frac{\D \mathbb{E}[f(x_t)]}{\D t} = \mathbb{E}[\mathscr{L}f(x_t)]$.

\section{Stochastic stability of quantum feedback control processes}

\subsection{Stochastic stability}
In control theory stabilization of nonlinear systems
is usually investigated using Lyapunov theory.
In the 1960s, the stochastic counterpart of Lyapunov theory 
\cite{Hasminskii-1980}
was developed by Has'minski\v\i\ and others.
To prove the central results of this article, we first have to define 
asymptotic stability of stochastic processes.

\begin{definition}
	Let $W_t$ be a Wiener process on the canonical Wiener space
	$(\Omega, \mathscr{F}, \mathbb{P})$, and let $x$ obey the
	Itô equation on $\mathbb{R}^n$,
	\begin{equation}
		\D x_t = b(x_t) \D t + \sigma(x_t) \D W_t,
		\label{eq-Ito}
	\end{equation}
	where $b$, $\sigma: \mathbb{R}^n \to \mathbb{R}^n$ satisfy the usual
	growth and Lip\-schitz conditions for existence and uniqueness of
	solutions \cite{Oeksendal-1998}.
	Then an equlibrium solution $x_*$ of Eq.~(\ref{eq-Ito}), 
	i.e., a solution satisfying $b(x_*) = \sigma(x_*) = 0$,
	is called \emph{stable in probability} if
	\begin{equation}
		\lim_{x_0 \to x_*} \mathbb{P} 
		\Big[ \sup_{t\geqq 0} |x_t - x_*| > \varepsilon \Big] = 0
		\qquad
		\forall \varepsilon > 0.
	\end{equation}
	It is called \emph{asymptotically stable} 
	if it is stable in probability and
	\begin{equation}
		\lim_{x_0 \to x_*} \mathbb{P} 
		\left[ \lim_{t\to \infty} |x_t - x_*| = 0 \right] = 1
		.
	\end{equation}
	It is called
	\emph{globally stable} if it is stable in probability and
	\begin{equation}
		\mathbb{P} 
		\left[ \lim_{t\to \infty} |x_t - x_*| = 0 \right] = 1
		.
	\end{equation}
	$x_*$ is called
	\emph{exponentially stable in $p$-th moment}, $p \in \mathbb{N}$,
	\cite{Higham-et-al-2003}
	if there exists a pair of constants $a$, $\alpha>0$ such that
	\begin{equation}
		\mathbb{E} 
		\big[ |x_t - x_*|^p \big] 
		\leqq
		a \, \mathbb{E}\big[ |x_0 - x_*|^p \big] \, \E^{-\alpha t}
	\end{equation}
	for all $t\geqq 0$.
	Especially for $p=1$, $x_*$ is then called
	\emph{exponentially stable in mean},
	and for $p=2$ \emph{exponentially stable in mean square}%
	.
	The smallest possible value of the 
	constant $a$ is referred to as the \emph{growth constant}, 
	and the largest possible value of
	$\alpha$ as the \emph{rate constant} or \emph{rate of convergence}.
\end{definition}
%
The first two notions are local properties, whereas the third one 
is a global property of the system.

\subsection{Imperfect detection efficiency}

If quantum feedback control is performed with only imperfect detection efficiency,
i.e., $0 < \eta < 1$, the following theorem yields a sufficient condition for
the global asymptotical stability of its final state.
Although a similar result for a special controller has been shown already in 
\cite[§\,IV.F]{van-Handel-et-al-2005} by numerical semialgebraic methods, 
here a general analytical criterion relating the controller parameters is given.

\begin{satz}
	\label{satz-asymptotically-stable}
	Consider 
	a quantum feedback control process 
	of a spin-$\frac{1}{2}$ quantum system, described by the
	density operators in the state space $D^2$
	and by the stochastic master equation (\ref{eq-SME-j=1/2}),
	with 
	the probe parameter measurement rate $M>0$,
	the detection efficiency $\eta$ $\in$ $(0,1]$
	and the controller
	\begin{equation}
		B(\lambda, \nu)
		= g_1 \lambda + g_2 \nu
		.
		\label{eq-controller}
	\end{equation}
	Assume that for the feedback gain parameters $g_1$ and $g_2$
	there exist real constants $c>1$, $0<d<2(c-1)$ such that
	the maximum $f_{\max}$ of
	the auxiliary function $f:[0,\frac12] \times [-\pi,\pi] \to \mathbb{R}$,
	\begin{eqnarray}
		f(r,\theta) 
		& \hspace*{-1.0ex} = \hspace*{-1.0ex} & \textstyle
		[M - (c-1)g_1] (1-\cos\theta)
		{}- dg_1 r \sin \theta (1-\cos\theta)
		\nonumber \\*
		& \hspace*{-1.0ex} \hspace*{-1.0ex} & \textstyle
		{}+ [d g_1 (1+\cos\theta)\, r - (c-1) g_2 - \frac {d(g_1+M)}{2}]
		\sin \theta
		\nonumber \\*
		& \hspace*{-1.0ex} \hspace*{-1.0ex} & \textstyle
		{}+ 2 d g_2 (1+\cos\theta) ( 2 r \cos\theta - \frac{1}{2})
		\nonumber \\*
		& \hspace*{-1.0ex} \hspace*{-1.0ex} & \textstyle
		{}- 4M\eta (1-\cos\theta) ((1+\cos\theta)\,r - \frac12)^2 
		\nonumber \\*
		& \hspace*{-1.0ex} \hspace*{-1.0ex} & \textstyle
		{}- 4M\eta (1+\cos\theta) ( (1+\cos\theta)\, r - 1)^2 
		,
		\label{condition-g1-g2}
	\end{eqnarray}
	is negative, i.e., $f_{\max} < 0$.
	Then the state $(0,0)\in D^2$, corresponding to
	$\rho_0$ $=$ $|1\rangle\langle 1|$,
	of the quantum system undergoing a quantum feedback control process 
	is globally stable. 
	Moreover, condition (\ref{condition-g1-g2}) implies
	\begin{equation}
		g_1 > \frac{(1-\eta)M}{c-1} \geqq 0,
		\qquad
		-\frac{4M\eta}{d} < g_2 < 0.
		\label{condition-g1-g2-2}
	\end{equation}
\end{satz}
\begin{beweis}	
	Defining the function
	\begin{equation}
		V(\lambda,\nu) = c \nu + d \lambda \nu - \lambda^2 - \nu^2,
		\label{eq-V}
	\end{equation}
	we have $V(0,0)=0$ and $V(\lambda,\nu) > 0$ for
	$(\lambda,\nu)\in D^2\setminus\{\fett{0}\}$.
	Since $0\leqq \lambda^2 \leqq \nu(1-\nu)$ 
	and $\lambda \geqq -\frac12$, we have
	$
		V=(c-1) \nu + d \lambda \nu - (\lambda^2 + \nu(\nu-1) )
		\geqq (c-1)\nu - \frac{d}{2} \nu
	$, i.e.,
	\begin{equation}
		\textstyle
		V(\lambda,\nu) \geqq (c-\frac{d}{2} -1)\, \nu,
		\label{eq-V>cnu}
	\end{equation}
	and especially $V>0$ for $\nu>0$.
	Moreover by (\ref{eq-controller}),
	\begin{eqnarray}
		\mathscr{L}V
		& \hspace*{-1.0ex} = \hspace*{-1.0ex} & \textstyle
		(g_1 \lambda + g_2 \nu)
		[(d(\nu^2 - \frac{\nu}{2} - \lambda^2) - (c-1)\lambda]
		\nonumber \\*
		& \hspace*{-1.0ex} \hspace*{-1.0ex} & \textstyle
		{}- \frac M2 \lambda (d \nu - 2\lambda)
		\nonumber \\*
		& \hspace*{-1.0ex} \hspace*{-1.0ex} & \textstyle
		{}- 4M\eta \left[\lambda^2 (\nu - \frac12)^2
		+ \nu^2 (\nu - 1)^2 
		\right]
		\nonumber \\* 
		& \hspace*{-1.0ex} = \hspace*{-1.0ex} & \textstyle
		[M - d  (g_1 \lambda + g_2 \nu) - (c-1) g_1] \lambda^2
		{}+ d g_2 \nu^2 (\nu - \frac12)
		\nonumber \\*
		& \hspace*{-1.0ex} \hspace*{-1.0ex} & \textstyle
		{}+ [d g_1 (\nu - \frac12) - (c-1) g_2 - \frac {dM}{2}] \lambda\nu
		\nonumber \\*
		& \hspace*{-1.0ex} \hspace*{-1.0ex} & \textstyle
		{}- 4M\eta \left[\lambda^2 (\nu - \frac12)^2
		+ \nu^2 (\nu - 1)^2 
		\right]
		.
		\label{eq-LV-1}
	\end{eqnarray}
	We see immediately that $\mathscr{L}V(0,0) = 0$.
	We will prove next that $\mathscr{L}V(\lambda,\nu) < 0$ for
	$(\lambda,\nu) \in D^2\setminus \{\fett{0}\}$.
	Using the coordinates $(r,\theta)$, where $r\in(0,\frac12]$ and
	$\theta \in (-\pi, \pi)$, given by
	$r=\frac{\lambda^2+\nu^2}{2\nu}$, $\tan\frac{\theta}{2} = \frac{\lambda}{\nu}$,
	we have
	$(\lambda,\nu)$ 
	$=$ $r(\sin\theta, 1+\cos\theta)$.
	Thus we obtain
	\begin{equation}
		f(r,\theta) 
		= \frac{\mathscr{L}V\big(\lambda(r,\theta), \nu(r,\theta)\big)}{r^2 (1+\cos\theta)}
		.
		\label{eq-LV-f}
	\end{equation}
	By the assumption of the Theorem, $f(r,\theta) < 0$ for
	$(r,\theta) \in [0,\frac12] \times (-\pi,\pi)$, hence
	$\mathscr{L}V(\lambda,\nu) < 0$ on $D^2\setminus \{\fett{0}\}$.
	Therefore,
	$V$ is a strict Lyapunov function on $D^2$ with the only asymptotically stable
	state $(\lambda,\nu)=0$. 

	Since $f(r,\pm\pi) = 2( (1-\eta)M - (c-1) g_1)$, 
	as well as
	$f(r,0) = -32M\eta (r^2 - (1+\frac{dg_2}{4M\eta})r + \frac{4M\eta + dg_2}{16 M\eta})$,
	assumption
	(\ref{condition-g1-g2}) implies (\ref{condition-g1-g2-2}).
\end{beweis}

Therefore, a quantum feedback control process satisfying the assumptions of
Theorem \ref{satz-asymptotically-stable} drives a spin-$\frac12$ quantum system
to the state $\binom{0}{0} \in D^2$,
{no matter in which quantum state the system is initially.}
This holds true even for the worst case, when the initial state is 
$\rho_i = |0\rangle \langle 0| = \binom{0}{1} \in D^2$. Note that a standard 
state reduction 
measurement would leave the quantum system in this state with certainty.

The next example shows that there indeed \emph{exist} 
parameter constellations satisfying Theorem \ref{satz-asymptotically-stable}.

\begin{beispiel}
	For the experimentally controlled parameters
	$g_1=\frac{3M}{4}$, $g_2=-\frac{M}{4}$, 
	and $\eta=\frac{1}{2}$, and 
\begin{figure}[htp]
\centering
	\begin{footnotesize}
	\unitlength=1ex
	\begin{picture}(32,27)
		\put(-5,-.5){\includegraphics[width=45ex]{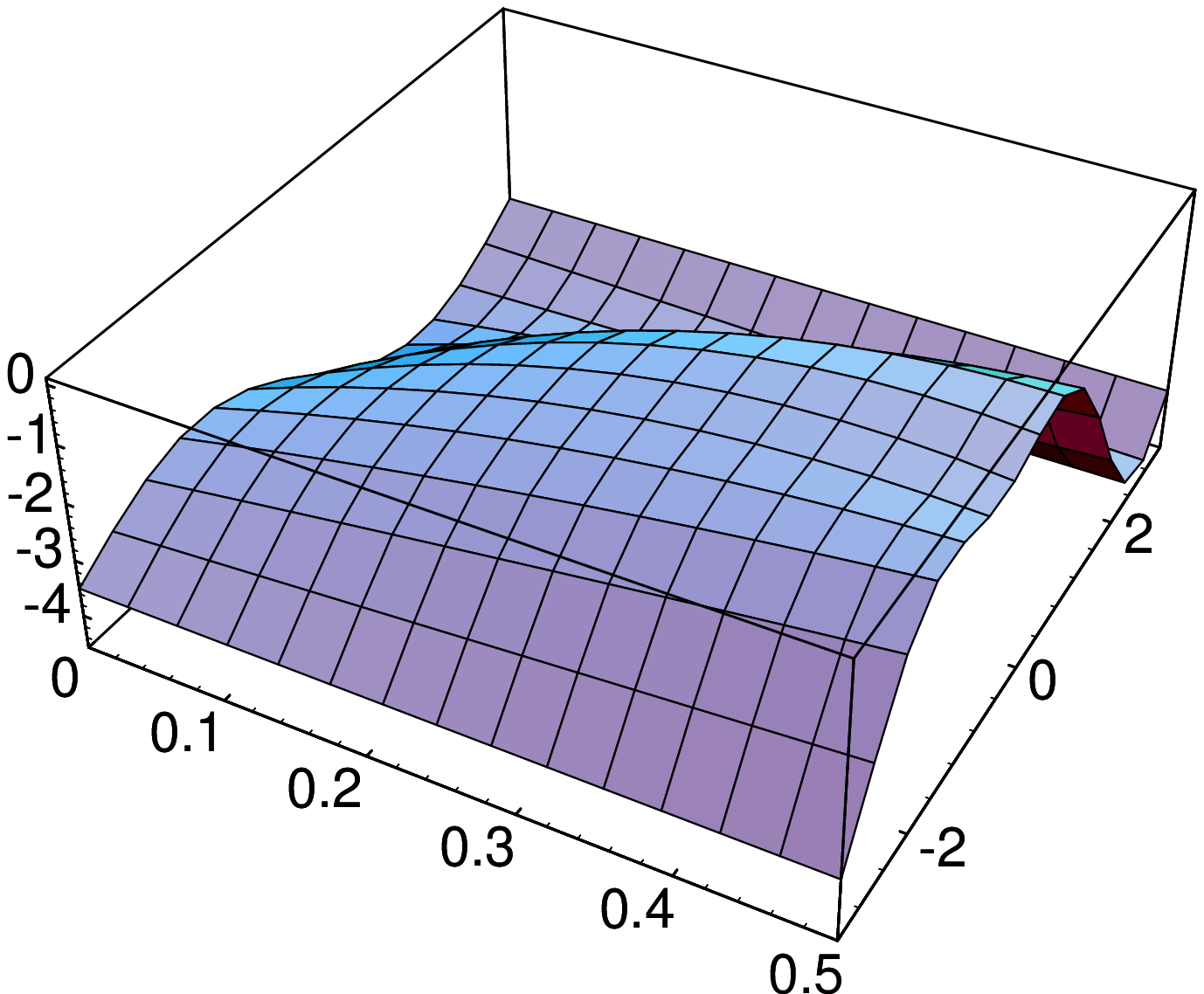}}
		\put( 0,12){\makebox(0,0)[r]{$f$}}
		\put(10,3){\makebox(0,0)[t]{$r$}}
		\put(30,7){\makebox(0,0)[l]{$\theta$}}
	\end{picture}
	\qquad
	\begin{picture}(28,27)
		\put(-8,-.5){\includegraphics[width=45ex]{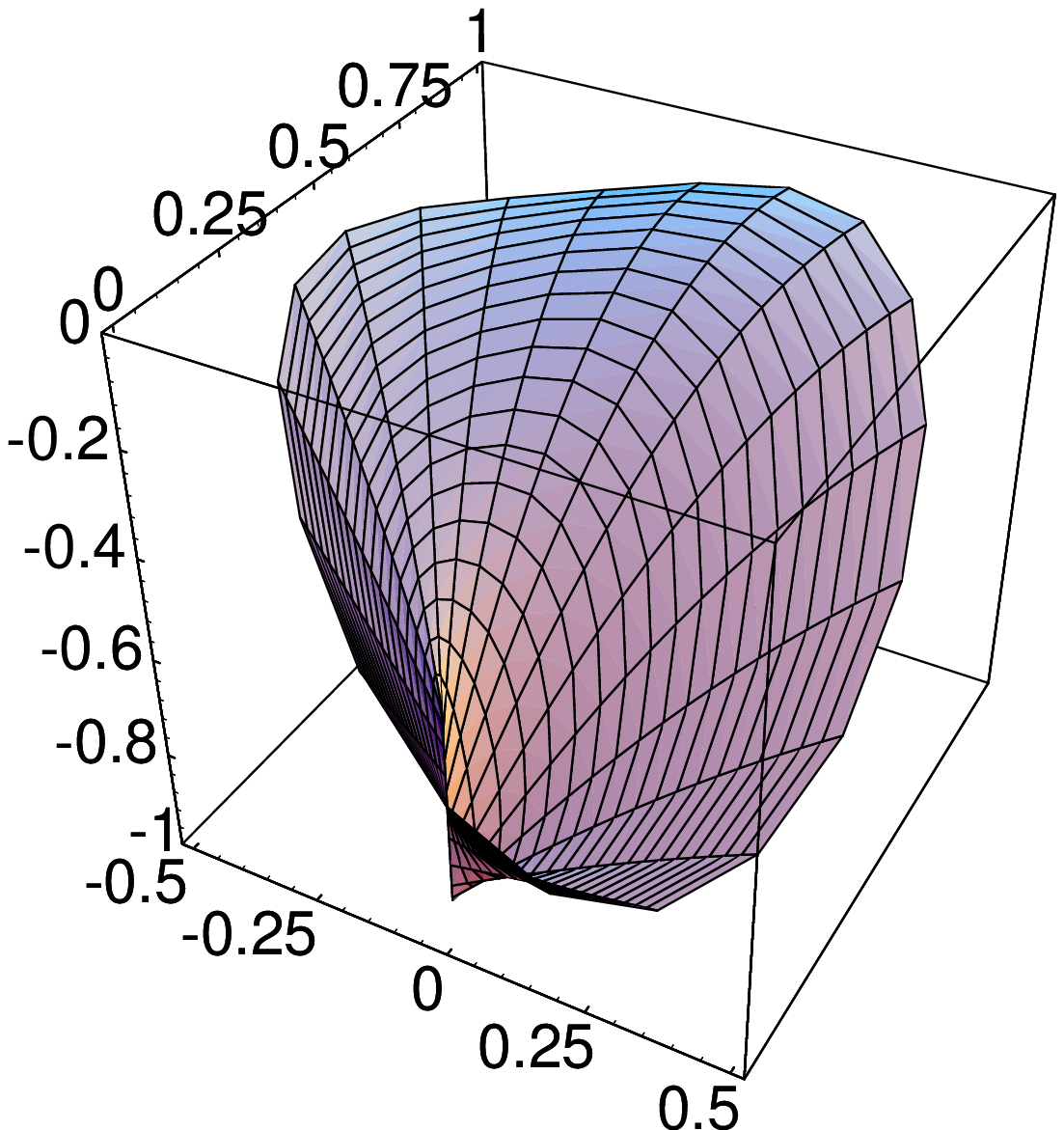}}
		\put( 6,25){\makebox(0,0)[r]{$\nu$}}
		\put( 1,12){\makebox(0,0)[r]{$z$}}
		\put(10,1){\makebox(0,0)[t]{$\lambda$}}
	\end{picture}
	\end{footnotesize}
	\caption{\label{fig-Lyapunov}
		(Color online)
		The auxiliary function $h$ in (\ref{condition-g1-g2})
		for $g_1=\frac{3M}{4}$, $g_2=-\frac{M}{4}$, 
		$\eta=\frac{M}{2}$,
		$c=4$, $d=2$. Left figure: The graph of $f(r,\theta)$.
		Right figure: The corresponding 2-manifold imbedded in
		3-space $(\lambda,\nu, z)$, parametrized by
		$\lambda=r\sin\theta$, $\nu=(1+\cos\theta)r$, $z=\frac{f(r,\theta)}{5}$
		for $0<r\leqq\frac12$, $|\theta| < \pi$.
	}
\end{figure}
	the constants $c=4$, $d=2$,
	the function $h$ in (\ref{condition-g1-g2})
	is negative on $D^2\setminus \{\fett{0}\}$, which may be seen
	graphically (Fig.~\ref{fig-Lyapunov}).
\end{beispiel}

\subsection{Pure states and perfect detection efficiency}
For pure quantum states and perfect detection efficiency, 
we are able to estimate the expected
running time for a quantum feedback control process more precisely, as is shown in the
following theorem.

\begin{satz}
	\label{satz-T-qfc}
	Consider a quantum feedback control process 
	of a spin-$\frac{1}{2}$ quantum system of pure quantum states, 
	described by the
	stochastic master equation (\ref{eq-SME-j=1/2})
	with 
	the probe parameter measurement rate $M>0$,
	the perfect detection efficiency $\eta=1$,
	and the controller $B(t)$ of (\ref{eq-controller})
	with feedback gain parameters
	\begin{equation}
		g_1 > 0, \quad g_2<0
		.
		\label{eq-cond-g1-g2}
	\end{equation} 	
	Then this process has an expected running time
	\begin{equation}
		T_{\mathrm{qfc}}(M,g_1)
		= O\big(\E^{- \left(g_1 + M \right) t/2}\big)
		,
		\label{eq-T-qfc}
	\end{equation}
	where $t$ denotes the time duration since the process start.
	In particular, $\fett{0} \in D^2$ is globally exponentially stable in mean.
\end{satz}
\begin{beweis}
	For a quantum feedback control process of a spin-$\frac12$ system, performed with
	perfect efficiency,
	the space of all density operators $\rho$ of pure states of the system can be reduced
	to the circle
	\begin{equation}
		S^1
		= \{ (\sin \theta, 1+\cos\theta) \in \mathbb{R}^2: \theta \in (-\pi,\pi]\}
	\end{equation}
	with center $(0,\frac12)$ and radius $\frac12$, 
	\cite[§IV.C]{van-Handel-et-al-2005},
	and the stochastic master equation (\ref{eq-SME})
	describing the conditional evolution of a single qubit, 
	with perfect efficiency $\eta=1$, then is
	reducible to the one-dimensional It{\^o} equation
	\begin{equation}
		\textstyle
		\D \theta_t
		= ( B(t) - \frac M2 \sin \theta_t \cos\theta_t) \D t
		- \sqrt M \sin\theta_t \D W_t
		.
		\label{eq-SME-j=1/2-pure}
	\end{equation}
	Choosing the controller as
	\begin{equation}
		B(t)
		= \frac{g_1}{2} \sin \theta_t + \frac{g_2}{2} (1+\cos\theta_t)
		,
	\end{equation}
	the system stabilizes the state $\theta=\pm\pi$, 
	which we mark as $|1\rangle \langle 1|$.
	Eq.~(\ref{eq-SME-j=1/2-pure}) has a unique solution on the
	interval $[-\pi,\pi]$
	because its coefficients satisfy the sufficient
	Lip\-schitz and growth conditions 
	\cite[Theor.\,5.2.1]{Oeksendal-1998}.
	Since the diffusion coefficient 
	\begin{equation}
		\sigma(\theta) = \sqrt{M} \sin\theta
	\end{equation}
	vanishes at $\theta=0$, 	
	we have to consider the two intervals 
	$J^- = (-\pi,0)$ and $J^+=(0,\pi)$ separately.
	Since moreover the drift coefficient
	\begin{equation}
		b(\theta) = B(\theta) - \frac{M}{2} \sin\theta \cos\theta
	\end{equation}
	satisfies $b(-\pi) = b(\pi) = 0$ and $b(0)=g_2<0$,
	the state $\theta=0$ is 
	a reflecting barrier (`entrance boundary')
	for states
	in $J^-$, but an absorbing barrier (`exit boundary')
	for states in $J^+$, whereas
	$\theta=\pm\pi$ both are absorbing barriers 
	\cite[§5.2.1]{Gardiner-1990}.
	
	To estimate the expected time that a given pure state
	requires to reach the desired state
	$|1\rangle\langle1|$, represented by $\theta=\pm\pi$,
	we have to compute the 
	expected first exit time $T$ 
	for the random variable $\theta$ to leave the interval
	$J^-$ or $J^+$, respectively.
	$T$ is the solution
	of the inhomogeneous linear differential equation
	\cite[§5.2.7]{Gardiner-1990}, \cite[§10.9]{Wilmott-1998}
	\begin{equation}
		\frac{\partial T}{\partial t}
		+ \frac{1}{2}\, \sigma^2(\theta)\, 
		\frac{\partial^2 T}{\partial \theta^2}
		+ b(\theta)\, \frac{\partial T}{\partial \theta}
		= -1
		\label{eq-first-exit-time}
	\end{equation}
	on $J^\pm$
	under the boundary conditions
	\begin{equation}
		T(\pm\pi,t) = T(0+,t)
		=
		\frac{\partial}{\partial \theta} T(0-,t) = 0,
		\quad
	\end{equation}
	and $T(\theta,0) = f(\theta)$.
	With the change of variable
	\begin{equation}
		x = \frac{1 + \cos\theta}{\sin\theta}
		  = \cot\frac{\theta}{2}
		\label{eq-x-theta}
	\end{equation}
	we obtain the relations $\sin\theta = \frac{2x}{1+x^2}$
	and $\cos \theta = \frac{x^2 - 1}{x^2 + 1}$, i.e.,
	$\frac{\sigma^2}{2} = \frac{2Mx^2}{(1+x^2)^2}$ and
	$b =\frac{2g_1 x}{1+x^2} + \frac{2 g_2 x^2}{1+x^2} - \frac{M(x-x^2)}{(1+x^2)^2}$.
	Moreover,
	$\D\theta = -\frac{2 \D x}{1 + x^2}$, i.e.,
	\begin{equation}
		\textstyle
		\frac{\partial}{\partial \theta} 
		= - \frac{1+x^2}{2}\frac{\partial}{\partial x},
		\quad
		\frac{\partial^2}{\partial \theta^2} 
		= \frac{(1+x^2)^2}{4}\frac{\partial^2}{\partial x^2}
		- \frac{x(1+x^2)}{2}\frac{\partial}{\partial x}
		.
	\end{equation}
	Hence Eq.~(\ref{eq-first-exit-time}) is rewritten as
	$\frac{\partial T}{\partial t} - \frac{M}{2} L T = -1$, where
	\begin{equation}
		L 
		= - x^2 \frac{\partial^2}{\partial x^2}
		+ h(x) 
		\frac{\partial}{\partial x}
	\end{equation}
	with
	\begin{equation}
		h(x) =
		\frac{g_1}{M} x + \frac{g_2}{M} x^2
		+ \frac{x - 3x^3}{1+x^2}
		,
	\end{equation}
	and where the boundary conditions $T(0,t) = 0$,
	$T(x,t) \to 0$ as $x \to \infty$, and
	$\frac{\partial}{\partial x} T(x,t) \to 0$ as $x \to -\infty$ hold.
	Since for $|x|<1$ we have $\frac{1}{1+x^2} = \sum_{0}^{\infty} (-1)^\nu x^{2\nu}$,
	we can write $h$ as
	\begin{equation}
		h(x) =
		\frac{g_1}{M} x + \frac{g_2}{M} x^2
		+ (x - 3x^3) \sum_{\nu=0}^\infty (-1)^\nu x^{2\nu}
		\label{eq-h-2}
	\end{equation}
	for $|x|<1$.
	An eigenvalue $\lambda$ of $L$ is given by the equation $Ly = \lambda y$.
	We use the ansatz $y(x) = \sum_0^\infty a_k x^k$.
	By the boundary conditions, $y(0)=0$, hence $a_0=0$.
	Moreover,
	$
		y'(x) = \sum_{1}^\infty k a_k x^{k-1},
	$
	and
	$
		y''(x) = \sum_{2}^\infty k(k-1) a_k x^{k-2},
	$
	hence 
	$$
		Ly(x) = \sum_{k=2}^\infty k(k-1) a_k x^{k} 
		+ h(x) \sum_{k=1}^\infty k a_k x^{k-1}
		.
	$$
	For $|x|<1$, 
	the series $\sum_{0}^\infty (-1)^\nu x^{2\nu}$ is absolutely convergent and
	we have 
	$
		(\sum_{1}^N k a_k x^{k-1}) (\sum_{0}^\infty (-1)^\nu x^{2\nu})
		= \sum_{1}^N c_k
	$
	with 
	$
	c_k
	= k a_k \sum_{\nu=0}^k (-1)^{k-\nu} x^{3k-2\nu-1}
	$,
	i.e., $c_1=a_1(1 - x^2)$, $c_2 = 2a_2 (x - x^3 + x^5)$,
	$c_3 = 3a_3 (x^2 - x^4 + x^6 - x^8)$.
	Hence 
	$Ly(x)$ has the following coefficients for the first powers of $x$,
	\begin{eqnarray}
		x 
		& \hspace*{-1ex} : &
		\left(\frac{g_1}{M} + 1\right) a_1 \\
		x^2 
		& \hspace*{-1ex} : &
		2 \left(\frac{g_1}{M} + 2\right) a_2 
		+ \frac{g_2}{M} a_1
		\\
		x^3 
		& \hspace*{-1ex} : &
		3 \left(\frac{g_1}{M} + 3\right) a_3
		+ 2 \cdot \frac{g_2}{M} a_2
		- 3 a_1
	\end{eqnarray}
	if $|x|<1$. Especially, the lowest power of $h(x)$ is the
	term $\left( \frac{g_1}{M} + 1 \right)$. 
	Let $y_{n}(x) = \sum_{n}^\infty a_k x^k$ for $n\in\mathbb{N}$,
	i.e., $a_k=0$ for $k \leqq n$.
	Then the lowest power of $Ly_n(x)$ is the term
	$\left( \frac{g_1}{M} + n \right) n a_{n} x^{n}$.
	Setting $Ly_n(x) = \lambda_n y_n(x)$ 
	and comparing the coefficients, we then get the eigenvalue
	\begin{equation}
		\lambda_n = \left( \frac{g_1}{M} + n \right) n
		\label{eq-eigenvalues}
	\end{equation}
	corresponding to the function $y_n$.
	In turn, once the eigenvalue is specified, the coefficients $a_{n+1}$, $a_{n+2}$, \ldots,
	are determined recursively by comparing the coefficients of $Ly_n$ and $\lambda_n y_n$.
	Although the above arguments hold true only for $|x|<1$, $\lambda_n$ is the
	eigenvalue corresponding to $y_n$ for the entire domains of definition,
	$x\in (-\infty,0)$ and $(0,\infty)$, respectively.
	Thus the smallest 
	eigenvalue of $L$ is $\lambda_1 = \frac{g_1}{M} + 1$.
	Now, $L$ can be expressed as $Ly =-\frac{1}{r}(py')'$ with
	$p(x) = \exp(-\int \frac{h(x)}{x^2}\D x)$, i.e.,
	\begin{equation}
		p(x)
		= \frac{1+x^2}{|x|^{\frac{g_1}{M}}} \E^{-\frac{g_2}{M}x}
		,
	\end{equation}
	and $r(x) = \frac{p(x)}{x^2}$,
	satisfying the boundary conditions 
	$y(0,t) = 0$,
	$y(x,t) \to 0$ as $x \to \infty$, 
	$\frac{\partial}{\partial x} y(x,t) \to 0$ as $x \to -\infty$, and 
	$y(x,0) = \frac{M}{2} f(x)$. Thus, we have a Sturm-Liouville problem on each
	interval $(-\infty,0)$ and $(0,\infty)$ separately, possessing the eigensolutions
	$y_n$ corresponding to the eigenvalues $\lambda_n$ of (\ref{eq-eigenvalues}),
	and our initial problem (\ref{eq-first-exit-time}) has the solutions
	\begin{eqnarray}
		T(\theta,t)
		& \hspace*{-1.0ex} = \hspace*{-1.0ex} &
		\int_{J^\pm} G(\theta,\alpha,t)\, r(\alpha) f(\alpha) \D \alpha
		\nonumber \\
		& \hspace*{-1.0ex}  \hspace*{-1.0ex} &
		-\int_0^t \int_{J^\pm} G(\theta,\alpha, t-\tau)\, r(\alpha) \D \alpha \D \tau
	\end{eqnarray}
	with
	\begin{equation}
		G(\theta,\alpha,t)
		=
		\sum_{n=1}^\infty \frac{y_n(\theta)y_n(\alpha)}{\|y_n\|^2} 
		\textstyle
		\E^{- \left(g_1 + Mn \right) n t/2}
	\end{equation}
	where $\theta=2 \arctan x$ and $\alpha = 2 \arctan x'$
	according to (\ref{eq-x-theta}), 
	and $\|y_n\|^2 = \int_{J^\pm} r y_n^2 \D x$
	.
	Hence an arbitrarily given initial state $\theta_0$, i.e.,
	$f(\theta) = \delta_{\theta_0}(\theta)$, is pushed exponentially fast in time $t$
	to one of the final states $\theta=\pm \pi$, or $x=0$, because
	\begin{eqnarray}
		T(\theta, t)
		& \hspace*{-.75ex} = \hspace*{-.75ex} &
		r(\theta_0)
		\sum_{n=1}^\infty \frac{y_n(\theta)y_n(\theta_0)}{\|y_n\|^2}
		\left(\frac{1+\lambda_n}{\lambda_n}\, \E^{-\lambda_n t} - \frac{1}{\lambda_n}\right)
		\nonumber \\
		& \hspace*{-.75ex} \leqq \hspace*{-.75ex} &
		2 \, \E^{-\lambda_1 t} \,
		r(\theta_0) 
		\sum_{n=1}^\infty \frac{y_n(\theta)y_n(\theta_0)}{\|y_n\|^2}
	\end{eqnarray}
	i.e., Eq.~(\ref{eq-T-qfc}),
	for the smallest eigenvalue $\lambda_1$ in (\ref{eq-eigenvalues}).
\end{beweis}

\begin{beispiel}
	Consider the the case
	$g_1=M$, $g_2=-\frac{M}{2}$ given in 
	\cite[Fig.~3\,(b)]{van-Handel-et-al-2005}.
	Then the eigenvalues $\mu$ of the
	expected time $T$ to set the quantum system into the state
	$|1\rangle \langle 1|$ given by (\ref{eq-eigenvalues})
	are bounded by $\mu \geqq \frac{M}{2} \lambda_1 = M$.
	Therefore, the greater the measurement rate $M$, the greater is
	the smallest possible eigenvalue of $T$.
\end{beispiel}


\section{Discussion}
In this article, a stability criterion for a
quantum feedback control process has been introduced,
as well as its expected running time in case of perfect detection efficiency.
In Theorem \ref{satz-asymptotically-stable},
a sufficient limit for the experimental
control parameters leading to globally stable quantum feedback control processes
acting on a spin-$\frac12$ quantum system are given. 
The proof consists of the application of Lyapunov's method to the stochastic
differential equation governing the quantum state evolution
under feedback control.

In Theorem \ref{satz-T-qfc} it is shown that,
for perfect detection efficiency, the quantum feedback control process
terminates even exponentially fast in time.
The proof bases on the power series ansatz 
$y=\sum_n^\infty a_k x^k$ for the
derived equation $\frac{\partial T}{\partial t} - \frac{M}{2} LT = -1$ determining
the expected time $T$, yielding eigenfunctions $y_n$ with corresponding
positive eigenvalues $\lambda_n$ by Eq.~(\ref{eq-eigenvalues}).
Mathematically, $T$ is the expected first exit time.
Theorem \ref{satz-T-qfc} implies that the expected
running time $T_{\mathrm{qfc}}$ of a quantum feedback control process
does not depend on the probability
neither of the desired state, nor of the initial state.
Former numerical investigations indicate
that the running time
of quantum feedback control algorithms, and thus 
$T_{\mathrm{qfc}}$,
is about a tenth of the decoherence time
\cite{Ahn-et-al-2002} up to the order of the decoherence time
\cite{Atkins-et-al-2005}.

Thus we are left with the unsatisfactory
situation that the general case of quantum feedback control with imperfect
detection efficiency could not yet be proved to be exponentially stable, in contrast
to the marginal case of pure states and perfect detection efficiency.
Of course, the fact that a Lyapunov function proving exponential stability
could not be found does not
mean that there does not exist any at all. However, by the proof of 
Theorem \ref{satz-asymptotically-stable} such a Lyapunov function
cannot be of the form (\ref{eq-V}),
since with
$\fett{x} = {\lambda \choose \nu}\in D^2$,
$|\fett{x}|^2 = \lambda^2 + \nu^2 = 2r\nu = 2r^2(1+\cos \theta),$
i.e., with Eq.~(\ref{eq-LV-f}),
\begin{equation}
	\textstyle
	\frac{f_{\min}}{2}\, |\fett{x}|^2
	\leqq
	\mathscr{L}V(\fett{x})
	\leqq
	\frac{f_{\max}}{2}\, |\fett{x}|^2
	\leqq 0
	,
	\label{eq-V-LV-1}
\end{equation}
but $V$ has a linear term such that there does not exist a constant $\alpha>0$
satisfying $\mathscr{L}V(\fett{x}) \leqq -\alpha V(\fett{x})$.
By Theorem \ref{satz-T-qfc}, however, a Lyapunov function may exist
satisfying this criterion at least for pure states.
Thus for future mathematical investigation the question remains:
Can quantum feedback control, including purification of a mixed state, 
be exponentially stable in general?






\end{document}